\documentclass[aps,prc,showpacs,twocolumn,nofootinbib,nobibnotes,groupedaddress]{revtex4-1}

\usepackage{calligra}
\usepackage{longtable}
\usepackage{graphicx}
\usepackage{dcolumn}
\usepackage{bm}
\usepackage{amsmath,amssymb}
\usepackage{longtable}

\newcommand{\Slash}[1]{\ooalign{\hfil/\hfil\crcr$#1$}}

\newcommand{\ket}[1]{| \, #1 \, \rangle}

\newcommand{\re}{\text{Re }}
\newcommand{\im}{\text{Im }}
\newcommand{\Tr}{\text{Tr }}

\usepackage[normalem]{ulem}  
\usepackage[dvips]{color} 

\renewcommand\sout{\bgroup \color{red} \ULdepth=-.5ex \ULset}

\begin{document}


\title{Nature of the $D_{0}$ meson in the $D\pi$ scattering with chiral symmetry}


\author{Takumi~Sugiura}
\affiliation{Department of Physics, Kyoto University,
Kyoto 606-8502, Japan}

\author{Tetsuo~Hyodo}
\email[]{hyodo@yukawa.kyoto-u.ac.jp}
\affiliation{Yukawa Institute for Theoretical Physics, Kyoto University, Kyoto 606-8502, Japan}


\date{\today}

\begin{abstract}
We study the nature of the scalar $D_{0}$ meson from the viewpoint of chiral symmetry. With the linear representation of chiral symmetry, we construct the $D\pi$ scattering amplitude satisfying the chiral low-energy theorem, in which the $D_{0}$ meson appears as an $s$-wave resonance. We show that the properties of the $D_{0}$ meson can be successfully reproduced as the chiral partner of the $D$ meson coupled with the $D\pi$ scattering states.  At the same time, we find that the spectral function and the pole position of $D_{0}$ are not very sensitive to the reduction of the chiral condensate, indicating the importance of the dressing of the bare state by the $D\pi$ molecular component.

\end{abstract}




\maketitle

\section{Introduction}


There have been a lot of newly observed hadrons~\cite{Tanabashi:2018oca} thanks to the recent developments of high-energy experiments. The findings of various exotic states suggest possible configurations of hadrons beyond the simple $\bar{q}q$ and $qqq$ classifications in conventional quark models. The suggestions for exotic configurations include, for instance, multiquarks, hadronic molecules, gluon hybrids, and so on. Investigations of understanding the nature of the exotic hadrons are intensively performed both experimentally and theoretically~\cite{Brambilla:2010cs,Hosaka:2016pey,Guo:2017jvc}.

Among many hadrons, charmed mesons composed of $c\bar{u}$ or $c\bar{d}$ are of particular interest from the viewpoint of QCD symmetries. In the massless limit of the light up ($u$) and down ($d$) quarks, QCD exhibits chiral SU(2)$_{L}\times $SU(2)$_{R}$ symmetry. Important low-energy properties of hadrons are constrained by chiral symmetry and its spontaneous breaking~\cite{Hosaka:2001ux,Scherer:2012xha}. In the limit of infinitely heavy mass of the charm ($c$) quark, hadrons with a single charm quark follows the heavy quark symmetry~\cite{Manohar:2000dt}, which also governs the dynamics of those hadrons. Thus, the heavy-light mesons provide a unique playground where two distinct symmetries in QCD interplay with each other. These symmetries are also important to study of the property change of the charmed mesons in nuclear matter~\cite{Hosaka:2016ypm}.

In this work, we focus on the scalar $D_{0}$ meson, which was firstly observed as a broad peak in the $D\pi$ invariant mass distribution in 2004~\cite{Link:2003bd}. Because $D_{0}$ has an opposite parity to the ground state pseudoscalar $D$ meson, it can be regarded as a chiral partner of the $D$ meson~\cite{Nowak:1992um,Bardeen:1993ae,Bardeen:2003kt,Suenaga:2017deu,Suenaga:2018kta}. On the other hand, its broad decay width to the $D\pi$ state implies the picture of the hadronic molecular state which is dynamically generated by the meson-meson interaction~\cite{Kolomeitsev:2003ac,Hofmann:2003je,Guo:2006fu,Gamermann:2006nm,Guo:2009ct,Altenbuchinger:2013vwa,Du:2017ttu,Du:2017zvv,Guo:2018tjx,Du:2019oki}. $D_{0}$ is also regarded as a flavor partner of the $D_{s0}(2317)$ state in the strangeness sector, which is known as a candidate of the exotic hadron.

The aim of this paper is to combine two pictures on the $D_{0}$ state, the chiral partner of $D$ and the hadronic molecule. To this end, we utilize the linear representation of chiral symmetry to incorporate the chiral partner structure, and describe the $D_{0}$ meson as a resonance in the scattering amplitude through  the nonperturbative resummation. This framework is analogous to the analysis of the $\sigma$ meson in the $\pi\pi$ scattering in Refs.~\cite{Basdevant:1969sz,Basdevant:1970nu,Jido:2000bw,Hyodo:2010jp}. The nature of the $D_{0}$ resonance is then studied by the property change with respect to the chiral condensate.

The paper is organized as follows. In Sec.~\ref{sec:form}, we start from the effective Lagrangian of charmed mesons, respecting chiral and heavy quark symmetries in QCD. The $D\pi$ scattering amplitude is then constructed, with the special emphasis of the chiral low-energy theorem. In Sec.~\ref{sec:numerical}, we numerically study the properties of the $D_{0}$ meson which is realized as a resonance in the $D\pi$ scattering amplitude. The last section is devoted to a summary of this work.

\section{Formulation}\label{sec:form}

\subsection{Effective Lagrangian}

To include the chiral partner structure of charmed mesons, we employ the effective Lagrangian based on the linear representation of chiral symmetry~\cite{Nowak:1992um,Bardeen:1993ae,Bardeen:2003kt,Suenaga:2017deu}. We first introduce the heavy-light meson fields $H_{L,R}\sim Q\bar{q}_{L,R}$, which transform as
\begin{align}
    H_{L,R}\to S H_{L,R}g_{L,R}^{\dag} ,
\end{align}
under chiral symmetry $(g_{L},g_{R})\in$ SU(2)$_{L}\times $SU(2)$_{R}$ and heavy quark spin symmetry $S\in$ SU(2)$_{S}$. The light mesons $\sigma$ and $\pi^{a}$ $(a=1,\ldots,3)$ are collected in the matrix field in the isospin space $M=\sigma+i\pi^{a}\tau^{a}\sim q_{L}\bar{q}_{R}$, which transforms as 
\begin{align}
    M\to g_{L} M g_{R}^{\dag} ,
\end{align}
under chiral SU(2)$_{L}\times $SU(2)$_{R}$ symmetry. The effective Lagrangian with no derivatives of the light meson fields, which is invariant under chiral and heavy quark spin symmetries, can be constructed as
\begin{align}
    \mathcal{L}
    &=-\Tr[H_{L}(iv\cdot \partial)\bar{H}_{L}]
    -\Tr[H_{R}(iv\cdot \partial)\bar{H}_{R}] \nonumber \\
    &\quad -\frac{\Delta_{m}}{2f_{\pi}}\Tr[H_{L}M\bar{H}_{R}
    +H_{R}M^{\dag}\bar{H}_{L}] 
    \label{eq:HLLagrangian} ,
\end{align}
where $\bar{H}_{L,R}\equiv \gamma^{0}H_{L,R}^{\dag}\gamma^{0}$, $v_{\mu}$ is the vector to specify the reference frame of the heavy quark, and the trace is taken over the Dirac space.\footnote{Note that $H_{L,R}$ ($\bar{H}_{L,R}$) is a row (column) vector in the isospin space, because it contains a light antiquark (quark).} In the interaction term, the coupling constant $\Delta_{m}$ is normalized by the pion decay constant $f_{\pi}$ for later convenience. To establish the connection to physical heavy-light mesons, we first introduce the negative (positive) parity heavy-light meson multiplet $H$ ($G$) as $H_{L,R}=[G\pm iH\gamma_{5}]/\sqrt{2}$. In the charm quark sector, $H$ ($G$) contains pseudoscalar $D$ and vector $D^{*}$ (scalar $D_{0}$ and axial vector $D_{1}$) as the heavy quark spin doublet, which are collected as 
\begin{align}
    H
    &=\frac{1+\Slash{v}}{2}
    (iD_{v}\gamma_{5}+\Slash{D}_{v}^{*}) , \label{eq:Hfield}\\
    G
    &=\frac{1+\Slash{v}}{2}
    (D_{0v}+i\gamma_{5}\Slash{D}_{1v}^{*}) .\label{eq:Gfield}
\end{align}
Substituting these into Eq.~\eqref{eq:HLLagrangian}, we obtain the effective Lagrangian of heavy-light mesons. The relevant terms for the present study are given by
\begin{align}
    \mathcal{L}
    &=\partial_{\mu}D\partial^{\mu}D^{\dag}
    -m^{2}DD^{\dag}
    +\partial_{\mu}D_{0}\partial^{\mu}D_{0}^{\dag}
    -m^{2}D_{0}D_{0}^{\dag} \nonumber \\
    &\quad +\frac{m\Delta_{m}}{2f_{\pi}}
    [D(M+M^{\dag})D^{\dag}-D_{0}(M+M^{\dag})D_{0}^{\dag} \nonumber \\
    &\quad -D_{0}(M-M^{\dag})D^{\dag}+D(M-M^{\dag})D_{0}^{\dag}]
    +\dotsb
    \label{eq:DLagrangian} ,
\end{align}
where $D=(D^{0},D^{+})$, $D^{\dag}=(\bar{D}^{0},D^{-})^{t}$ and we have adopted the relativistic notation as in Ref.~\cite{Suenaga:2017deu} with $m$ being the chiral invariant mass of the charmed mesons. The first line of Eq.~\eqref{eq:DLagrangian} denotes the free fields of $D$ and $D_{0}$, having the common mass $m$ due to the chiral partner structure, and the second and third lines represent the interactions with $\sigma$ and $\pi$, respectively. It is a consequence of chiral symmetry that the couplings to $\sigma$ and $\pi$ are dictated by a common coupling constant $\Delta_{m}$.

The light meson part of the effective Lagrangian is 
\begin{align}
    \mathcal{L}
    &=\frac{1}{4}\Tr\Bigl[\partial^{\mu}M\partial_{\mu}M^{\dag}
    -m_{0}^{2}MM^{\dag}\nonumber\\
    &\quad -\frac{\lambda}{4}(MM^{\dag})^{2}
    +\epsilon(M+M^{\dag})\Bigr] 
    \label{eq:LightLagrangian} ,
\end{align}
where the trace is taken over the isospin space. For a negative $m_{0}^{2}$, chiral symmetry is spontaneously broken at the mean field level. Denoting the mean field value of the $\sigma$ field as $\bar{\sigma}$ and redefining the $\sigma$ field as the fluctuation around $\bar{\sigma}$, we can read off the hadron masses from Eqs.~\eqref{eq:DLagrangian} and \eqref{eq:LightLagrangian} as
\begin{align}
    m_{\pi}^{2}
    &=\frac{\epsilon}{\bar{\sigma}}=m_{0}^{2}+\lambda\bar{\sigma}^{2}, 
    \label{eq:pimass}\\
    m_{\sigma}^{2}
    &=m_{0}^{2}+3\lambda\bar{\sigma}^{2}, 
    \label{eq:sigmamass}\\
    M_{D}
    &= m-\frac{\Delta_{m}\bar{\sigma}}{2f_{\pi}} , \label{eq:Dmass}\\
    M_{D_{0}}
    &= m+\frac{\Delta_{m}\bar{\sigma}}{2f_{\pi}} .\label{eq:D0mass}
\end{align}
In the light meson sector, three parameters in the Lagrangian $m_{0}$, $\lambda$, and $\epsilon$ are determined by the physical values of $m_{\pi}$, $m_{\sigma}$, and the chiral condensate in vacuum, which corresponds to the pion decay constant $\bar{\sigma}=f_{\pi}$. Therefore, the mass difference between $D$ and $D_{0}$ in vacuum is given by $\Delta_{m}=M_{D_{0}}-M_{D}$. 

As mentioned above, the parameter $\Delta_{m}$ also determines the $D_{0}D\pi$ vertex in the third line of Eq.~\eqref{eq:DLagrangian}. Thus, for a given $M_{D_{0}}$, we can predict the decay width $\Gamma_{D_{0}}$ in the perturbative calculation. However, if we use the central value adopted by the Particle Data Group (PDG)~\cite{Tanabashi:2018oca} $M_{D_{0}}=2318$ MeV to determine $\Delta_{m}$, we obtain $\Gamma_{D_{0}}=1115$ MeV, which largely deviates from the PDG value $\Gamma_{D_{0}}=267\pm 40$ MeV. In other words, $D_{0}$ as a pure chiral partner state is not consistent with the experimental data.\footnote{See also the recent discussion on the PDG value in Ref.~\cite{Du:2019oki}.} In fact, huge decay width obtained by the perturbative estimate indicates the strong coupling, for which the nonperturbative treatment of the coupling is needed. In the following, we regard $D_{0}$ field in the Lagrangian as a bare state, and describe the physical $D_{0}$ state as a resonance in the nonperturbative $D\pi$ scattering amplitude.

\subsection{Tree-level $D\pi$ interaction}

\begin{figure*}[tbp]
    \centering
    \includegraphics[width=15cm,bb=0 0 692 92]{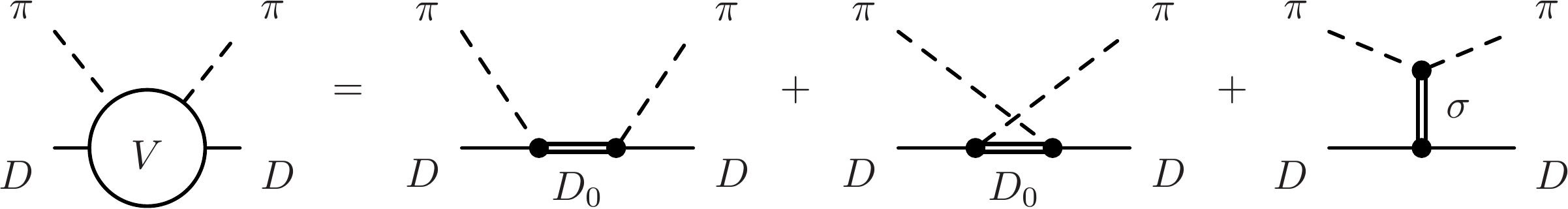}
    \caption{\label{fig:treediagrams}
    Feynman diagrams for the tree-level $D\pi$ scattering. Solid, dashed, and double lines represent the pion, $D$, and $D_{0}$ (sigma) propagators, respectively.}
\end{figure*}

Let us construct the $D\pi$ scattering amplitude at tree level. From Eq.~\eqref{eq:DLagrangian}, we have the $DD\sigma$ and $D_{0}D\pi$ vertices
\begin{align}
    \mathcal{L}
    &=\frac{m\Delta_{m}}{f_{\pi}}D\sigma D^{\dag} 
    +\frac{im\Delta_{m}}{f_{\pi}}
    [D\pi^{a}\tau^{a}D_{0}^{\dag}-D_{0}\pi^{a}\tau^{a}D^{\dag}]
    \label{eq:Dlight} ,
\end{align}
and from Eq.~\eqref{eq:LightLagrangian}, we have $\sigma\pi\pi$ vertex
\begin{align}
    \mathcal{L}
    &=-\frac{m_{\sigma}^{2}-m_{\pi}^{2}}{2\bar{\sigma}}
    \sigma\pi^{a}\pi^{a}
    \label{eq:sigmapipi} .
\end{align}
From these vertices, the $D\pi$ scattering amplitude at tree level can be constructed by the $s$-channel $D_{0}$ exchange, $u$-channel $D_{0}$ exchange, and $t$-channel $\sigma$ exchange, as shown in Fig.~\ref{fig:treediagrams}.\footnote{In this work, we consider the single-channel $s$-wave $D\pi$ scattering, and the couplings to $s$-wave $D_{0}\sigma$ channel and $p$-wave $D^{*}\sigma$, $D_{1}\pi$ channels are not included. These channels are in higher energies and do not contribute to the chiral low-energy theorem.} Straightforward calculation leads to the tree-level amplitude in the particle basis. For instance, in the total charge $Q=+1$ sector, we obtain
\begin{align}
    V_{ij}(s,t,u)
    &=
    \begin{pmatrix}
    1 & -\sqrt{2} \\
    -\sqrt{2} & 2
    \end{pmatrix}
    \frac{m^{2}\Delta_{m}^{2}}{f_{\pi}^{2}}\frac{1}{s-M_{D_{0}}^{2}} \nonumber \\
    &\quad 
    +
    \begin{pmatrix}
    1 & \sqrt{2} \\
    \sqrt{2} & 0
    \end{pmatrix}
\frac{m^{2}\Delta_{m}^{2}}{f_{\pi}^{2}}\frac{1}{u-M_{D_{0}}^{2}} \nonumber \\
    &\quad 
    +\begin{pmatrix}
    -1 & 0 \\
    0 & -1
    \end{pmatrix}\frac{m\Delta_{m}(m_{\sigma}^{2}-m_{\pi}^{2})}{f_{\pi}\bar{\sigma}}\frac{1}{t-m_{\sigma}^{2}} ,
    \label{eq:Vcharge}
\end{align}
with $\pi^{0}D^{+}$ and $\pi^{+}D^{0}$ for $i=1$ and $2$, respectively. This coupled-channel interaction can be diagonalized by taking the isospin basis where the interaction $V^{I}$ for isospin $I$ reads
\begin{align}
    V^{1/2}(s,t,u)
    &=
    \frac{3(M_{D_{0}}^{2}-M_{D}^{2})^{2}}{4\bar{\sigma}^{2}}
    \frac{1}{s-M_{D_{0}}^{2}} \nonumber \\
    &\quad -\frac{(M_{D_{0}}^{2}-M_{D}^{2})^{2}}{4\bar{\sigma}^{2}}
    \frac{1}{u-M_{D_{0}}^{2}} \nonumber \\
    &\quad 
    -\frac{(M_{D_{0}}^{2}-M_{D}^{2})(m_{\sigma}^{2}-m_{\pi}^{2})}{2\bar{\sigma}^{2}}
    \frac{1}{t-m_{\sigma}^{2}}
    ,\label{eq:V1o2} \\
    V^{3/2}(s,t,u)
    &=
    \frac{(M_{D_{0}}^{2}-M_{D}^{2})^{2}}{2\bar{\sigma}^{2}}\frac{1}{u-M_{D_{0}}^{2}} \nonumber \\
    &\quad -\frac{(M_{D_{0}}^{2}-M_{D}^{2})(m_{\sigma}^{2}-m_{\pi}^{2})}{2\bar{\sigma}^{2}}
    \frac{1}{t-m_{\sigma}^{2}} ,
    \label{eq:V3o2}
\end{align}
with 
\begin{align}
    \begin{pmatrix}
    \ket{D\pi(I=1/2)} \\
    \ket{D\pi(I=3/2)}
    \end{pmatrix}
    &=
    \begin{pmatrix}
    \sqrt{\frac{1}{3}} & -\sqrt{\frac{2}{3}} \\
    -\sqrt{\frac{2}{3}} & -\sqrt{\frac{1}{3}}
    \end{pmatrix}
    \begin{pmatrix}
    \ket{\pi^{0}D^{+}} \\
    \ket{\pi^{+}D^{0}}
    \end{pmatrix} .
\end{align}
In Eqs.~\eqref{eq:V1o2} and \eqref{eq:V3o2}, we replace the coupling strength $m\Delta_{m}$ by the masses of the $D$ mesons using Eqs.~\eqref{eq:Dmass} and \eqref{eq:D0mass}. Because of the isospin symmetry of the effective Lagrangians, there is no transition between $I=1/2$ and 3/2 sectors. Note also that the $s$-channel pole term only exists in the $I=1/2$ sector, reflecting the isospin of the $D_{0}$ state.

\subsection{Chiral low-energy theorem}\label{subsec:LET}

Before proceeding to the calculation of the nonperturbative scattering amplitude, it is instructive to examine the results~\eqref{eq:V1o2} and \eqref{eq:V3o2} from the viewpoint of chiral low-energy theorem. Regarding pions as Nambu-Goldstone (NG) bosons associated with the spontaneous breaking of chiral symmetry, we can impose several conditions on the low-energy dynamics of pions. In particular, low-energy scattering of pions with any target hadron is determined only by the pion decay constant and the total isospin of the system, as a consequence of the Weinberg-Tomozawa theorem~\cite{Weinberg:1966kf,Tomozawa:1966jm}. In the case of the $D\pi$ scattering, the low-energy scattering should behave as
\begin{align}
    V^{I}
    &\propto 
    \frac{1}{\bar{\sigma}^{2}}[I(I+1)-11/4] \nonumber \\
    &=\frac{1}{\bar{\sigma}^{2}}\times 
    \begin{cases}
    -2 & (I=1/2) \\
    1 & (I=3/2)
    \end{cases} .
    \label{eq:WTth}
\end{align}
This is an important constraint on the hadron scattering amplitude in the following sense. First, from Eq.~\eqref{eq:WTth} the sign (either attractive or repulsive) and strength of the interaction are uniquely determined, according to the total isospin of the system. Second, the relation is insensitive to the internal structure of the target hadron, except for the isospin of the state. This means that, for instance, the $\pi N$ system also follows the same relation~\eqref{eq:WTth}, which was confirmed experimentally to a good accuracy by the $\pi N$ scattering lengths~\cite{Scherer:2012xha}. In chiral perturbation theory, the Weinberg-Tomozawa theorem is automatically satisfied by the leading order calculation, while a careful examination is needed in the linear representation as in the present study.

To study the low-energy (near-threshold) behaviors of Eqs.~\eqref{eq:V1o2} and \eqref{eq:V3o2}, we assign the meson momenta as $D(k)\pi(p) \to D(k^{\prime})\pi(p^{\prime})$. We regard the three-momentum of the mesons in the center-of-mass frame $|\bm{p}|=|\bm{k}|=|\bm{p}^{\prime}|=|\bm{k}^{\prime}|$ as a small quantity. Denoting it as $\mathcal{O}(Q)$, we expand the interaction $V^{I}$ in powers of $Q$. Masses of non-NG bosons, $m_{\sigma}$, $M_{D}$, and $M_{D_{0}}$ are counted as $\mathcal{O}(Q^{0})$, while the pion mass is regarded as $m_{\pi}\sim \mathcal{O}(Q^{1})$ according to the chiral counting. In this case, the energy of pions is counted as $\mathcal{O}(Q^{1})$ that we define $p_{0}=p_{0}^{\prime}\equiv \omega $. The energy of the $D$ mesons $k_{0}=k_{0}^{\prime}$ is counted as $M_{D}+\mathcal{O}(Q^{2})$. Then the Mandelstam variables are expanded as
\begin{align*}
    s
    &=M_{D}^{2}+2M_{D}\omega +\mathcal{O}(Q^{2}) ,\\ 
    t
    &=\mathcal{O}(Q^{2}) ,\\
    u
    &=M_{D}^{2}-2M_{D}\omega +\mathcal{O}(Q^{2}) .
\end{align*}
We finally obtain the low-energy behaviors of Eqs.~\eqref{eq:V1o2} and \eqref{eq:V3o2} as
\begin{align}
    V^{1/2}
    &=
    \frac{3}{4}
    \frac{M_{D_{0}}^{2}-M_{D}^{2}}{\bar{\sigma}^{2}}
    \left[-1-\frac{2M_{D}\omega }{M_{D_{0}}^{2}-M_{D}^{2}} \right] \nonumber \\
    &\quad -
    \frac{1}{4}\frac{M_{D_{0}}^{2}-M_{D}^{2}}{\bar{\sigma}^{2}}
    \left[-1+\frac{2M_{D}\omega }{M_{D_{0}}^{2}-M_{D}^{2}} \right] \nonumber \\
    &\quad 
    +\frac{2}{4}\frac{M_{D_{0}}^{2}-M_{D}^{2}}{\bar{\sigma}^{2}}
    +\mathcal{O}(Q^{2}) \nonumber \\
    &= -\frac{2M_{D}\omega }{\bar{\sigma}^{2}}
    +\mathcal{O}(Q^{2}) ,
    \label{eq:V1o2low}\\
    V^{3/2}
    &=
    \frac{M_{D}\omega }{\bar{\sigma}^{2}}
    +\mathcal{O}(Q^{2}) 
    \label{eq:V3o2low} .
\end{align}
We confirm that the results satisfy the relation~\eqref{eq:WTth} by the Weinberg-Tomozawa theorem. Furthermore, the energy dependence agrees with the prediction by chiral perturbation theory $V\propto (s-u)/4=M_{D}\omega +\mathcal{O}(Q^{2})$~\cite{Guo:2006fu}.

It is worth noting that the contribution from each channel has an $\mathcal{O}(Q^{0})$ term, in contradiction to the low-energy theorem which requires $\mathcal{O}(Q^{1})$ as the leading order. In other words, the use of the $s$-channel pole term with energy independent coupling is not consistent with the chiral low-energy theorem. By summing up all the channels required by the chiral invariant Lagrangian, the $\mathcal{O}(Q^{0})$ terms cancel out with each other, and we obtain $V^{I}\sim \mathcal{O}(Q^{1})$ in agreement with the low-energy theorem. This is analogous to the $\pi N$ scattering in the linear sigma model, where the sum of the $s$- and $u$-channel Born terms with pseudoscalar coupling leads to $\mathcal{O}(Q^{0})$ contribution, and the inclusion of the $t$-channel $\sigma$ exchange recovers the result of the Weinberg-Tomozawa theorem. 

\subsection{Nonperturbative $D\pi$ scattering amplitude}

To describe the $D_{0}$ meson as a resonance in the $D\pi$ scattering, we project the interaction $V(s,t,u)$ to $s$ wave, and perform nonperturbative resummation to satisfy the unitarity. From now on, we concentrate on the $I=1/2$ channel and suppress the isospin index. In the center of mass frame, the $t$ and $u$ variables can be specified by the total energy $\sqrt{s}$ and the scattering angle $\cos\theta=\bm{p}\cdot\bm{p}^{\prime}/|\bm{p}\cdot\bm{p}^{\prime}|$ with the momentum assignment $D(k)\pi(p) \to D(k^{\prime})\pi(p^{\prime})$. The $s$-wave part $V(\sqrt{s})$ is obtained by averaging over the scattering angle as
\begin{align}
    V(\sqrt{s})
    =
    \frac{1}{2}\int_{-1}^{1}d\cos\theta \ V(\sqrt{s},\cos\theta) .
    \label{eq:sprojection}
\end{align}
Defining the magnitude of the three momentum as 
\begin{align}
    p&\equiv 
    |\bm{p}|
    =|\bm{p}^{\prime}|
    =\frac{\lambda^{1/2}(s,m_{\pi}^{2},M_{D}^{2})}{2\sqrt{s}} ,
\end{align}
with the K\"all\'en function $\lambda(x,y,z)=x^{2}+y^{2}+z^{2}-2xy-2yz-2zx$, we obtain the expression for the $s$-wave projected interaction:
\begin{align}
    V(\sqrt{s})
    &= \frac{3(M_{D_{0}}^{2}-M_{D}^{2})^{2}}{4\bar{\sigma}^{2}}
    \frac{1}{s-M_{D_{0}}^{2}} \nonumber \\
    &\quad -\frac{(M_{D_{0}}^{2}-M_{D}^{2})^{2}}{4\bar{\sigma}^{2}}
    \frac{1}{4p^{2}}
    \ln\frac{M_{D_{0}}^{2}-\frac{(m_{\pi}^{2}-M_{D}^{2})^{2}}{s}}{M_{D_{0}}^{2}-\frac{(m_{\pi}^{2}-M_{D}^{2})^{2}}{s}+4p^{2}} \nonumber \\
    &\quad 
    -\frac{(M_{D_{0}}^{2}-M_{D}^{2})(m_{\sigma}^{2}-m_{\pi}^{2})}{2\bar{\sigma}^{2}}
    \frac{1}{4p^{2}}
    \ln\frac{m_{\sigma}^{2}}{m_{\sigma}^{2}+4p^{2}} .
\end{align}

Because the tree-level interaction does not satisfy the unitarity condition of the $S$ matrix, various resummation (unitarization) schemes have been proposed. Here we employ the N/D method~\cite{Oller:1998zr,Hyodo:2010jp}. By neglecting the contribution from the left-hand cut, we obtain the general expression of the scattering amplitude $T(\sqrt{s})$ from the tree-level interaction $V(\sqrt{s})$ as
\begin{align}
    T(\sqrt{s})
    &=\frac{1}{[V(\sqrt{s})]^{-1}-G(\sqrt{s})}
    \label{eq:Tmatrix}
\end{align}
with
\begin{align}
    G(\sqrt{s})
    &= \frac{1}{16\pi^{2}}
    \Biggl[a(\mu)
    +\ln\frac{m_{\pi}M_{D}}{\mu^{2}}
    +\frac{M_{D}^{2}-m_{\pi}^{2}}{2s}\ln\frac{M_{D}^{2}}{m_{\pi}^{2}} \nonumber  \\
    &\quad 
    +\frac{p}{\sqrt{s}}
    [\ln(s
    -m_{\pi}^{2}
    +M_{D}^{2}+2\sqrt{s}\; p)\nonumber \\
    &\quad 
    +\ln(s
    +m_{\pi}^{2}
    -M_{D}^{2}+2\sqrt{s}\; p) \nonumber \\
    &\quad 
    -\ln(-s
    +m_{\pi}^{2}
    -M_{D}^{2}+2\sqrt{s}\; p)\nonumber \\
    &\quad 
    -\ln(-s
    -m_{\pi}^{2}
    +M_{D}^{2}+2\sqrt{s}\; p) ] 
    \Biggr] ,
    \label{eq:Gfn}
\end{align}
where $a(\mu)$ is the subtraction constant at the subtraction scale $\mu$. This $G(\sqrt{s})$ function is obtained by the once-subtracted dispersion integral on the unitarity cut, and is essentially the same with the $D\pi$ loop function with the dimensional regularization. With this identification, Eq.~\eqref{eq:Tmatrix} corresponds to the resummation of the loop diagrams in the Bethe-Salpeter equation. 

A resonance in the $D\pi$ scattering is expressed as the pole of the analytically continued scattering amplitude in the second Riemann sheet of the complex energy plane. When the amplitude has a pole at $\sqrt{s}=z\in \mathbb{C}$, the mass $M_{R}$ and width $\Gamma_{R}$ of the resonance can be identified by
\begin{align}
    M_{R}
    &=\re z, \quad
    \Gamma_{R}
    = -2\ \im z .
    \label{eq:MassWidth}
\end{align}
Because of the nonperturbative resummation, the bare mass $M_{D_{0}}$ in the Lagrangian is shifted to $M_{R}-i\Gamma_{R}/2$, including the decay width to the $D\pi$ channel. Therefore, this resummation procedure can also be regarded as the renormalization of the $D_{0}$ propagator with the Dyson equation. We however note that the self-energy should include not only the simple $D\pi$ loop, as was done in Ref.~\cite{Suenaga:2017deu}, but also the crossed diagrams, in order to satisfy the chiral low-energy theorem.

As we discussed in Sec.~\ref{subsec:LET}, the low-energy limit of our interaction $V(\sqrt{s})$ reduces to the Weinberg-Tomozawa term, which can dynamically generate a $D\pi$ molecule state  without the explicit bare state~\cite{Kolomeitsev:2003ac,Hofmann:2003je,Guo:2006fu,Gamermann:2006nm,Guo:2009ct,Altenbuchinger:2013vwa,Du:2017ttu,Du:2017zvv,Guo:2018tjx,Du:2019oki}. On top of that, our interaction contains the explicit coupling to the bare $D_{0}$ state which is introduced as a chiral partner of $D$. Thus, our model is capable of describing both the chiral partner state and the $D\pi$ molecule state. In the following, we calibrate the model by empirical data and study the nature of the $D_{0}$ resonance in the $D\pi$ scattering amplitude.

\section{Numerical analysis}\label{sec:numerical}

\subsection{$D_{0}$ resonance}

Now we numerically study the property of the $D_{0}$ resonance. We first set the parameters of the model to the empirical values,
\begin{align}
    m_{\pi}&=138 \text{ MeV},
    \quad M_{D}=1867 \text{ MeV}, \\
    \bar{\sigma}&= 92.4 \text{ MeV},\quad
    m_{\sigma}= 550\text{ MeV} .
\end{align}
In addition, we need to specify one degree of freedom to determine the ultraviolet cutoff in Eq.~\eqref{eq:Gfn}. Here we set $\mu=1000$ MeV and use the subtraction constant $a$ at this scale as a free parameter for the cutoff degree of freedom. The bare mass $M_{D_{0}}$ in the Lagrangian is also used as a free parameter. 

As a physical input to determine the free parameters, we use the mass and width of $D_{0}$ in PDG~\cite{Tanabashi:2018oca}. According to Eq.~\eqref{eq:MassWidth}, we should have a pole of the $D\pi$ scattering amplitude at
\begin{align}
    \sqrt{s}&=2318\pm 29-i(134\pm 20) \text{ MeV},
    \label{eq:PDG}
\end{align}
Note however that the PDG values are obtained by the Breit-Wigner parametrization
which is recently challenged by Ref.~\cite{Du:2019oki}. At present, we simply adopt Eq.~\eqref{eq:PDG} as a representative value, but it is also possible to use the updated pole positions as suggested in Ref.~\cite{Du:2019oki}. Such analysis is reserved for a future work.

\begin{figure}[tbp]
    \centering
    \includegraphics[width=8cm,bb=0 0 514 370]{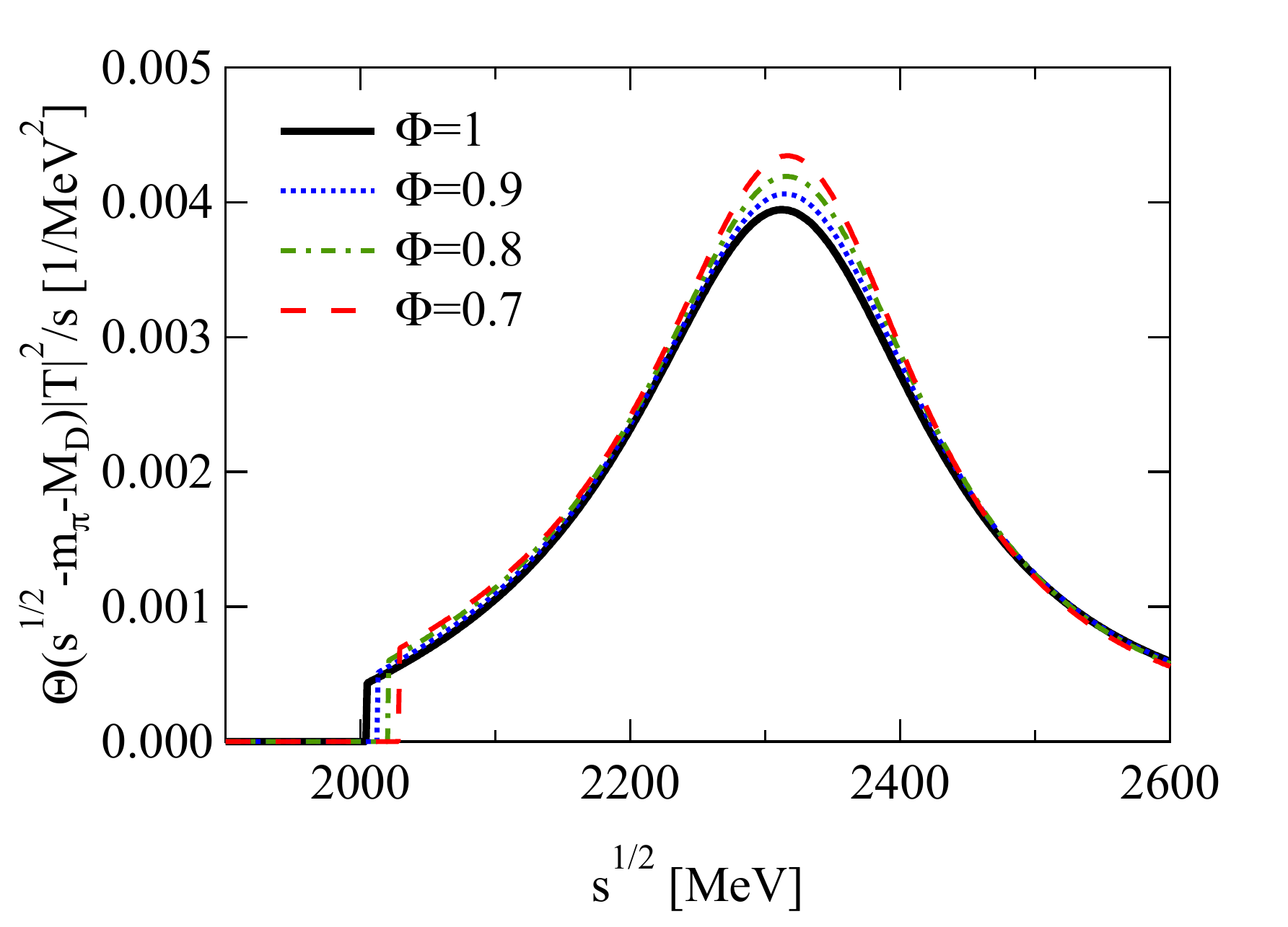}
    \caption{\label{fig:cross}
    Reduced $D\pi$ cross section $\Theta(\sqrt{s}-m_{\pi}-M_{D})|T(\sqrt{s})|^{2}/s$ in isospin $I=1/2$ channel. Solid line corresponds to the result with $\bar{\sigma}=f_{\pi}$, and the other lines show the results with modifying the chiral condensate $\bar{\sigma}=\Phi f_{\pi}$ by the reduction factor $\Phi$.}
\end{figure}

The best fit values of the bare mass and the subtraction constant are found to be $M_{D_{0}}=2024$ MeV and $a=2.5$, with which we obtain the $D_{0}$ pole at
\begin{align}
    \sqrt{s}&=2318-i135 \text{ MeV},
\end{align}
in good agreement with Eq.~\eqref{eq:PDG}. We also find a virtual state pole at $\sqrt{s}=1795$ MeV. Although the location of the pole is far below the threshold (2005 MeV), this may correspond to the one found in the sigma propagator in Refs.~\cite{Hidaka:2003mm,Hidaka:2004ht}. We show the reduced $D\pi$ cross section $\Theta(\sqrt{s}-m_{\pi}-M_{D})|T(\sqrt{s})|^{2}/s$ by solid line in Fig.~\ref{fig:cross}. This corresponds to the spectrum of the scalar channel with isospin $I=1/2$, and the peak of the $D_{0}$ resonance is seen. We also find that the pole position of Eq.~\eqref{eq:PDG} cannot be reproduced with some values of the bare mass, even if we vary the subtraction constant freely. Namely, the bare mass should lie in the region $2001\text{ MeV} \leq M_{D_{0}}\leq 2045$ MeV in order to satisfy Eq.~\eqref{eq:PDG} within the uncertainty.

\subsection{Reduction of chiral condensate}

While the chiral partner state is included in the model as a bare state, the bare mass of 2024 MeV is more than 300 MeV away from the actual pole position. This may indicate that the bare pole contribution is not very relevant to the structure of the physical state. To check this qualitatively, we vary the value of the chiral condensate $\bar{\sigma}$ from its vacuum value, and study the response of the spectrum and the pole position. This is partly motivated by the reduction of the chiral condensate in nuclear medium~\cite{Hayano:2008vn}, but the quantitative discussion on the in-medium spectrum will require the actual many-body calculation as was done in Refs.~\cite{Suenaga:2017deu,Suenaga:2018kta}.

Here we introduce a reduction parameter $\Phi$ as
\begin{align}
    \bar{\sigma}&=\Phi f_{\pi}
\end{align}
and vary it within $0.7\leq \Phi\leq 1$. The masses of the $D$ mesons are given by Eqs.~\eqref{eq:Dmass} and \eqref{eq:D0mass}, with fixed $m$ and $\Delta_{m}$. Namely, the mass of $D$ (bare mass of $D_{0}$) increases (decreases) as we decrease $\Phi$ from unity. For the light mesons, we assume that the pion mass is independent of $\Phi$ and the mass of sigma varies as $m_{\sigma}^{2}=m_{\pi}^{2}+2\lambda \bar{\sigma}^{2}$ with $\lambda$ fixed, as suggested in Ref.~\cite{Hyodo:2010jp}. Namely, the mass of sigma decreases along with the reduction of $\Phi$.

\begin{figure}[tbp]
    \centering
    \includegraphics[width=8cm,bb=0 0 394 282]{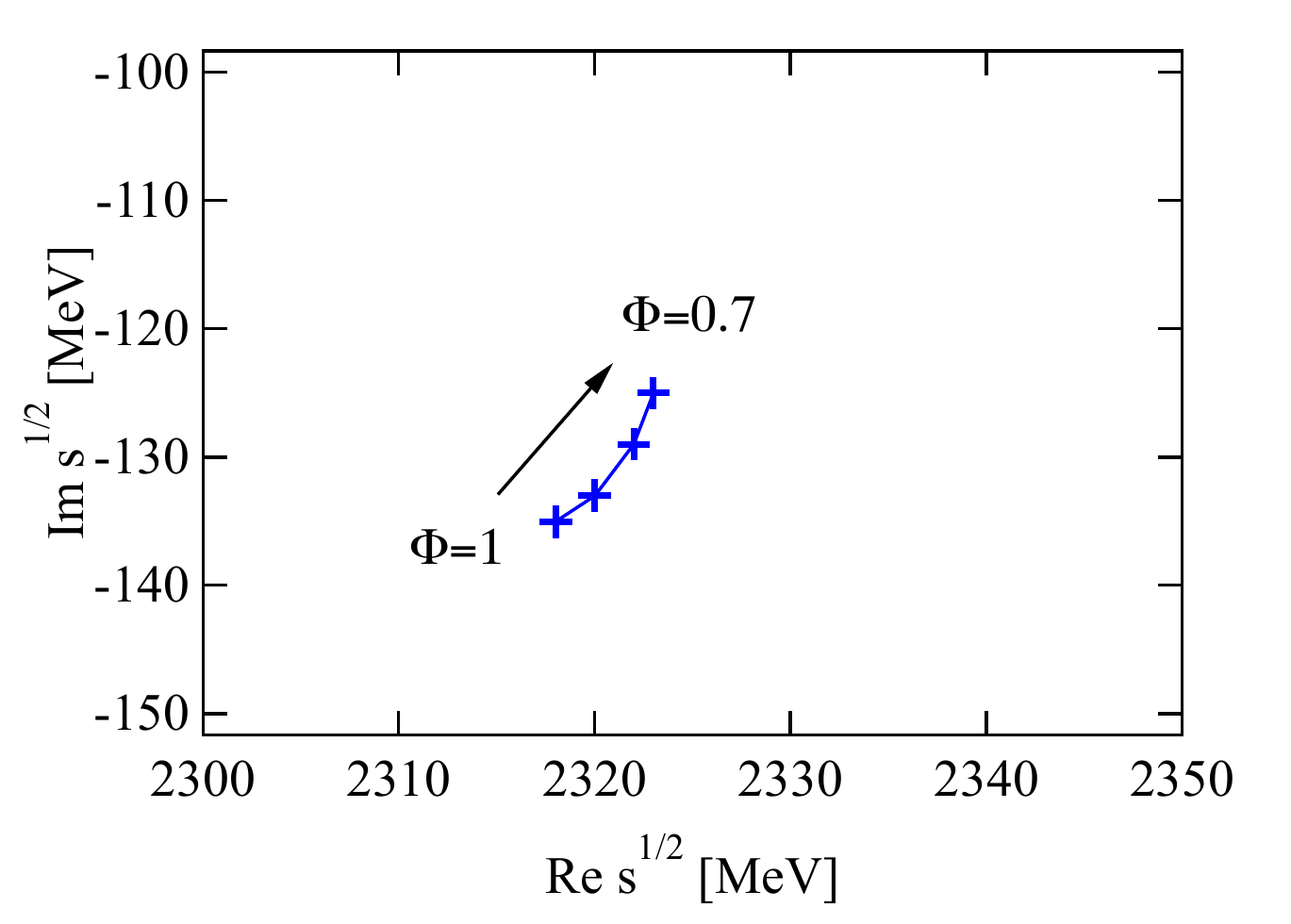}
    \caption{\label{fig:pole}
    Trajectory of the $D_{0}$ pole along with the reduction of chiral condensate. The arrow indicates the direction of the pole movement.}
\end{figure}

The modification of the $D_{0}$ spectrum along with the reduction of $\bar{\sigma}$ is shown in Fig.~\ref{fig:cross} by the dashed lines. We find that the spectrum is barely changed, even if the chiral condensate is reduced by 30 \%. This shows that the resonance is not sensitive to the change of the chiral condensate. This can also be seen by the trajectory of the pole shown in Fig.~\ref{fig:pole}. The change of the pole position is a few tens of MeV with the $30$ \% reduction of $\bar{\sigma}$. Moreover, the real part of the pole increases when the chiral condensate is reduced, although the bare mass is moved to the lower energies. We also note that the virtual state pole at 1795 MeV moves toward the threshold.  This is in line with the findings in Refs.~\cite{Hidaka:2003mm,Hidaka:2004ht}, although the pole locates still far away from the threshold ($\sqrt{s}=1864$ MeV at $\Phi=0.7$). 

As found in Ref.~\cite{Hyodo:2010jp}, the pole of the chiral partner state is naively expected to move toward the threshold with reducing the imaginary part due to the decrease of the phase space. We confirm this behavior by artificially tuning the subtraction constant ($a=0$) to have a narrow $D_{0}$ resonance (pole at $\sqrt{s}=2069-32i$ MeV with $\Phi=1$). This indicates that, in the physical $D_{0}$ resonance (with $a=2.5$), the dressing by the $D\pi$ molecule component is so large that it washes out the chiral partner nature of the bare state. In this way, we conclude that the $D\pi$ molecule component is important for the physical $D_{0}$ meson, on top of the bare state which is the chiral partner of the $D$ meson. 

\section{Summary}\label{sec:summary}

We have studied the properties of the $D_{0}$ resonance in the $D\pi$ scattering, utilizing the chiral effective Lagrangian. We introduce the chiral partner of the $D$ meson as an explicit bare state in the linear representation of chiral symmetry. At the same time, we respect the chiral low-energy theorem which leads to the dynamical generation of the $D\pi$ molecule state. Because the empirical data can be explained by this mode, we demonstrate that the physical $D_{0}$ resonance can be described as a superposition of the chiral partner of $D$ and the $D\pi$ molecular state. By studying the response to the modification of the chiral condensate, we find that the bare $D_{0}$ state is highly dressed by the $D\pi$ molecular component to realize the physical $D_{0}$ resonance. Thus, we conclude that the $D\pi$ molecular component dominates the internal structure of the $D_{0}$ meson.

As future prospects, we can improve the empirical information of the $D_{0}$ state as suggested by Ref.~\cite{Du:2019oki}. It is possible to use the updated pole position as an input, or to directly analyze the $B$ meson decay data. The generalization to three-flavor case is also important, because the thresholds of $D\eta$ and $DK$ channels are located around 2400 MeV, whose contribution may also affect to the $D_{0}$ resonance. In this case, the $D_{s0}(2317)$ can also be studied in the same framework. Another direction would be to explore the modification of the $D_{0}$ property in nuclear medium, which can be pursued with the techniques developed in Refs.~\cite{Suenaga:2017deu,Suenaga:2018kta}.

\section*{Acknowledgments}

The authors thank Masayasu Harada, Atsushi Hosaka, Feng-Kun Guo, Daisuke Jido, Daiki Suenaga, and Shigehiro Yasui for fruitful discussions. This work is supported in part by JSPS KAKENHI Grant No. 24740152, and by the Yukawa International Program for Quark-Hadron Sciences (YIPQS).



\begin{thebibliography}{10}

\bibitem{Tanabashi:2018oca}
M.~Tanabashi {\em et~al.},
\newblock Phys. Rev. D {\bf 98}, 030001 (2018).

\bibitem{Brambilla:2010cs}
N.~Brambilla {\em et~al.},
\newblock Eur. Phys. J. C {\bf 71}, 1534 (2011).

\bibitem{Hosaka:2016pey}
A.~Hosaka, T.~Iijima, K.~Miyabayashi, Y.~Sakai and S.~Yasui,
\newblock PTEP {\bf 2016}, 062C01 (2016).

\bibitem{Guo:2017jvc}
F.-K. Guo {\em et~al.},
\newblock Rev. Mod. Phys. {\bf 90}, 015004 (2018).

\bibitem{Hosaka:2001ux}
A.~Hosaka and H.~Toki,
\newblock {\em Quarks, baryons and chiral symmetry} (World Scientific,
  Singapore, 2001).

\bibitem{Scherer:2012xha}
S.~Scherer and M.~R. Schindler,
\newblock Lect. Notes Phys. {\bf 830}, pp.1 (2012).

\bibitem{Manohar:2000dt}
A.~V. Manohar and M.~B. Wise,
\newblock {\em Heavy quark physics} (Cambridge University Press, Cambridge,
  2000).

\bibitem{Hosaka:2016ypm}
A.~Hosaka, T.~Hyodo, K.~Sudoh, Y.~Yamaguchi and S.~Yasui,
\newblock Prog. Part. Nucl. Phys. {\bf 96}, 88 (2017).

\bibitem{Link:2003bd}
FOCUS, J.~M. Link {\em et~al.},
\newblock Phys. Lett. {\bf B586}, 11 (2004).

\bibitem{Nowak:1992um}
M.~A. Nowak, M.~Rho and I.~Zahed,
\newblock Phys. Rev. D {\bf 48}, 4370 (1993).

\bibitem{Bardeen:1993ae}
W.~A. Bardeen and C.~T. Hill,
\newblock Phys. Rev. D {\bf 49}, 409 (1994).

\bibitem{Bardeen:2003kt}
W.~A. Bardeen, E.~J. Eichten and C.~T. Hill,
\newblock Phys. Rev. D {\bf 68}, 054024 (2003).

\bibitem{Suenaga:2017deu}
D.~Suenaga, S.~Yasui and M.~Harada,
\newblock Phys. Rev. C {\bf 96}, 015204 (2017).

\bibitem{Suenaga:2018kta}
D.~Suenaga,
\newblock arXiv:1805.01709 [nucl-th].

\bibitem{Kolomeitsev:2003ac}
E.~E. Kolomeitsev and M.~F.~M. Lutz,
\newblock Phys. Lett. B {\bf 582}, 39 (2004).

\bibitem{Hofmann:2003je}
J.~Hofmann and M.~F.~M. Lutz,
\newblock Nucl. Phys. {\bf A733}, 142 (2004).

\bibitem{Guo:2006fu}
F.-K. Guo, P.-N. Shen, H.-C. Chiang, R.-G. Ping and B.-S. Zou
\newblock Phys. Lett. B {\bf 641}, 278 (2006).

\bibitem{Gamermann:2006nm}
D.~Gamermann, E.~Oset, D.~Strottman and M.~J. Vicente~Vacas,
\newblock Phys. Rev. D {\bf 76}, 074016 (2007).

\bibitem{Guo:2009ct}
F.-K. Guo, C.~Hanhart and U.-G. Meissner,
\newblock Eur. Phys. J. {\bf A40}, 171 (2009).

\bibitem{Altenbuchinger:2013vwa}
M.~Altenbuchinger, L.~S. Geng and W.~Weise,
\newblock Phys. Rev. D {\bf 89}, 014026 (2014).

\bibitem{Du:2017ttu}
M.-L. Du, F.-K. Guo, U.-G. Meissner and D.-L. Yao,
\newblock Eur. Phys. J. {\bf C77}, 728 (2017).

\bibitem{Du:2017zvv}
M.-L. Du {\em et~al.},
\newblock Phys. Rev. D {\bf 98}, 094018 (2018).

\bibitem{Guo:2018tjx}
Z.-H. Guo, L.~Liu, U.-G. Meissner, J.~A. Oller and A.~Rusetsky,
\newblock Eur. Phys. J. {\bf C79}, 13 (2019).

\bibitem{Du:2019oki}
M.-L. Du, F.-K. Guo and U.-G. Meissner,
\newblock arXiv:1903.08516 [hep-ph].

\bibitem{Basdevant:1969sz}
J.~L. Basdevant and B.~W. Lee,
\newblock Phys. Lett. B {\bf 29}, 437 (1969).

\bibitem{Basdevant:1970nu}
J.~I. Basdevant and B.~W. Lee,
\newblock Phys. Rev. D {\bf 2}, 1680 (1970).

\bibitem{Jido:2000bw}
D.~Jido, T.~Hatsuda and T.~Kunihiro,
\newblock Phys. Rev. D {\bf 63}, 011901 (2001).

\bibitem{Hyodo:2010jp}
T.~Hyodo, D.~Jido and T.~Kunihiro,
\newblock Nucl. Phys. A {\bf 848}, 341 (2010).

\bibitem{Weinberg:1966kf}
S.~Weinberg,
\newblock Phys. Rev. Lett. {\bf 17}, 616 (1966).

\bibitem{Tomozawa:1966jm}
Y.~Tomozawa,
\newblock Nuovo Cim. A {\bf 46}, 707 (1966).

\bibitem{Oller:1998zr}
J.~A. Oller and E.~Oset,
\newblock Phys. Rev. D {\bf 60}, 074023 (1999).

\bibitem{Hidaka:2003mm}
Y.~Hidaka, O.~Morimatsu, T.~Nishikawa and M.~Ohtani,
\newblock Phys. Rev. D {\bf 68}, 111901 (2003).

\bibitem{Hidaka:2004ht}
Y.~Hidaka, O.~Morimatsu, T.~Nishikawa and M.~Ohtani,
\newblock Phys. Rev. D {\bf 70}, 076001 (2004).

\bibitem{Hayano:2008vn}
R.~S. Hayano and T.~Hatsuda,
\newblock Rev. Mod. Phys. {\bf 82}, 2949 (2010).

\end{thebibliography}

\end{document}